\documentclass[12pt]{article}
\usepackage{graphicx, amssymb}
\graphicspath{{figs/}}

\usepackage[colorlinks=true,urlcolor=blue,
anchorcolor=blue,citecolor=blue,filecolor=blue,
linkcolor=blue,menucolor=blue,
linktocpage=true,pdfa=true]{hyperref}

\newcommand{\fig}[2]{\includegraphics[width=#1\textwidth]{#2}}
\usepackage[square,sort&compress,comma,numbers]{natbib}

\begin{document}
\title{Reconstruction of Energy of Ultra-High-Energy Cosmic Rays
Registered with a Fluorescence Telescope:\\ One Time Frame Might Be
Enough}
\author{M. Zotov (1) and A. Trusov (2) for the JEM-EUSO collaboration\\
\small
(1) Skobeltsyn Institute of Nuclear Physics, Lomonosov Moscow State
University\\
\small
(2) Faculty of Physics, Lomonosov Moscow State University}
\date{October 13, 2025}
\maketitle
\begin{abstract}

	We address the challenge of reconstructing the energy of three
	ultra-high-energy cosmic rays registered with a small fluorescence
	telescope EUSO-TA that operated in 2015 at the site of the Telescope
	Array experiment in Utah, USA. Each of these events was recorded
	within one time frame.  Conventional methods of energy reconstruction
	are not applicable in this case because the events do not have light
	curves but a single data point.  As an alternative, we consider a
	number of approaches based on artificial neural networks. We
	demonstrate that a signal recorded by a fluorescence telescope
	within one time frame might be enough to reconstruct energy of a
	primary particle with reasonable accuracy using an ensemble of simple
	convolutional neural networks.  Contrary to the conventional
	approach, reconstruction of the shower geometry is not needed for
	this.  More than this, preliminary estimates can be obtained even
	without recognizing the shower track.  However, there remain some
	problems that do not allow us to claim that the suggested method is
	universal and always works.  We discuss difficulties that we faced
	and possible ways of improving the method.

\end{abstract}

\section{Introduction}

Energy of primary particles is one of the key parameters to be inferred
from observations of cosmic rays (CRs), both direct and indirect.  At
energies above approximately 0.1~EeV, fluorescence telescopes play an
important role in CR studies~\cite{auger-fd, heat, lhaaso}.  This
happens because they provide an opportunity to directly observe the
development of extensive air showers (EASs) in the atmosphere via
registering their fluorescence light during clear moonless nights.  This
makes estimates based on these observations less dependent on models of
hadronic interactions than those obtained with surface detectors, which,
in its turn, suggests using them for calibration of other instruments.

Reconstruction of energy of a primary particle on the basis of
information collected by a ground-based fluorescence telescope (FT)
consists of the following major steps~\cite{flyseye-1985,kuempel-2008,
auger-fd}. First, the track of the shower on the photodetector (or a
group of photodetectors) must be recognized as a sequence of triggered
(activated) pixels. Next, one has to find out the shower geometry, which
usually consists of the determination of the shower-detector plane and
the consequent reconstruction of the shower axis within the plane.
Finally, one employs data about the light collected at the aperture as a
function of time to estimate energy of a primary particle. A similar
procedure was developed for orbital FTs~\cite{jemeuso-angular,
jemeuso-energy}.  It is important to stress that in both cases
reconstruction of energy is performed by fitting data points that
represent light detected in activated pixels with a \textit{curve} as a
function of time. That is why fluorescence detectors have time
resolution high enough to record light curves of extensive air showers.

Earlier we presented simple artificial neural networks (ANNs) aimed at
reconstruction of energy and arrival directions of ultra-high energy
cosmic rays (UHECRs) registered with two small fluorescence telescopes
developed by the JEM-EUSO collaboration: a stratospheric FT
EUSO-SPB2~\cite{spb2} and a ground-based EUSO-TA
telescope~\cite{eusota-2018}.  The method consisted of two main
steps~\cite{me}. First, the track of an extensive air shower registered
by a FT was recognised by a convolutional encoder-decoder
(CED)~\cite{segnet}. Next, parameters of a primary particle were
reconstructed by a six-layer convolutional neural network (CNN).  No
intermediate reconstruction of the shower geometry was needed.  The
pipeline demonstrated decent accuracy on simulated data. The CNNs could
be trained to reconstruct energy and arrival directions of UHECRs
independently or simultaneously.  In particular, this means that one
does not need to determine the geometry of an EAS in order to estimate
the energy of its primary particle with the help of the CNNs.

In what follows, we apply the method suggested in~\cite{me} to
reconstruct energy of three of the UHECRs registered by EUSO-TA in 2015
during its first test run at the site of the Telescope Array experiment.
We describe in detail the whole procedure and discuss problems that we
faced. We also outline some possible directions for the further
development of the method.  Results presented below are strongly based
on paper~\cite{me}, thus we shall denote it as Paper~I in what follows.

\section{EUSO-TA}

EUSO-TA is a small fluorescence telescope built as a prototype for
testing the design, electronics, software and other aspects of the
future orbital missions~\cite{eusota-2015, eusota-2018, eusota-2024}.
It operated at the site of the Telescope Array experiment in Utah, USA,
near its Black Rock Mesa FTs. EUSO-TA is equipped with two Fresnel
lenses of a diameter of 1~m and a concave focal surface, which consists
of $48\times48$ pixels.  The total field of view of the telescope equals
approximately $10.6^\circ\times10.6^\circ$, with the field of view of
one pixel equal to $0.2^\circ\times0.2^\circ$. During the campaign of
2015, the time resolution of the detector was equal to 2.5~$\mu$s.
The telescope was tested at several elevation angles: $10^\circ$,
$15^\circ$, $20^\circ$, $21^\circ$, and $25^\circ$.
The trigger was provided by the FTs at the Black Rock Mesa site.

Being a test-bed for larger orbital detectors, EUSO-TA was not
well-suited for ground observations of UHECRs. On the one hand, due to
its narrow field of view, it could only detect fluorescence light from
small portions of EASs (except for some highly inclined showers pointed
towards the aperture). On the other hand, its electronics was designed
for registering fluorescence light produced by EASs from the orbit
height of the International Space Station, around 400~km. For
comparison, the time resolution of FTs of the Pierre Auger Observatory
and the Telescope Array is equal to 100~ns~\cite{auger-fd, ta-fd}. As a
result, all UHECRs registered by EUSO-TA during the campaign of 2015
consisted of one or, at most, two active frames, so that one could not
obtain a light curve and reconstruct the energy of these events
following the conventional procedure.

\begin{figure}[!ht]
	\centerline{\fig{.25}{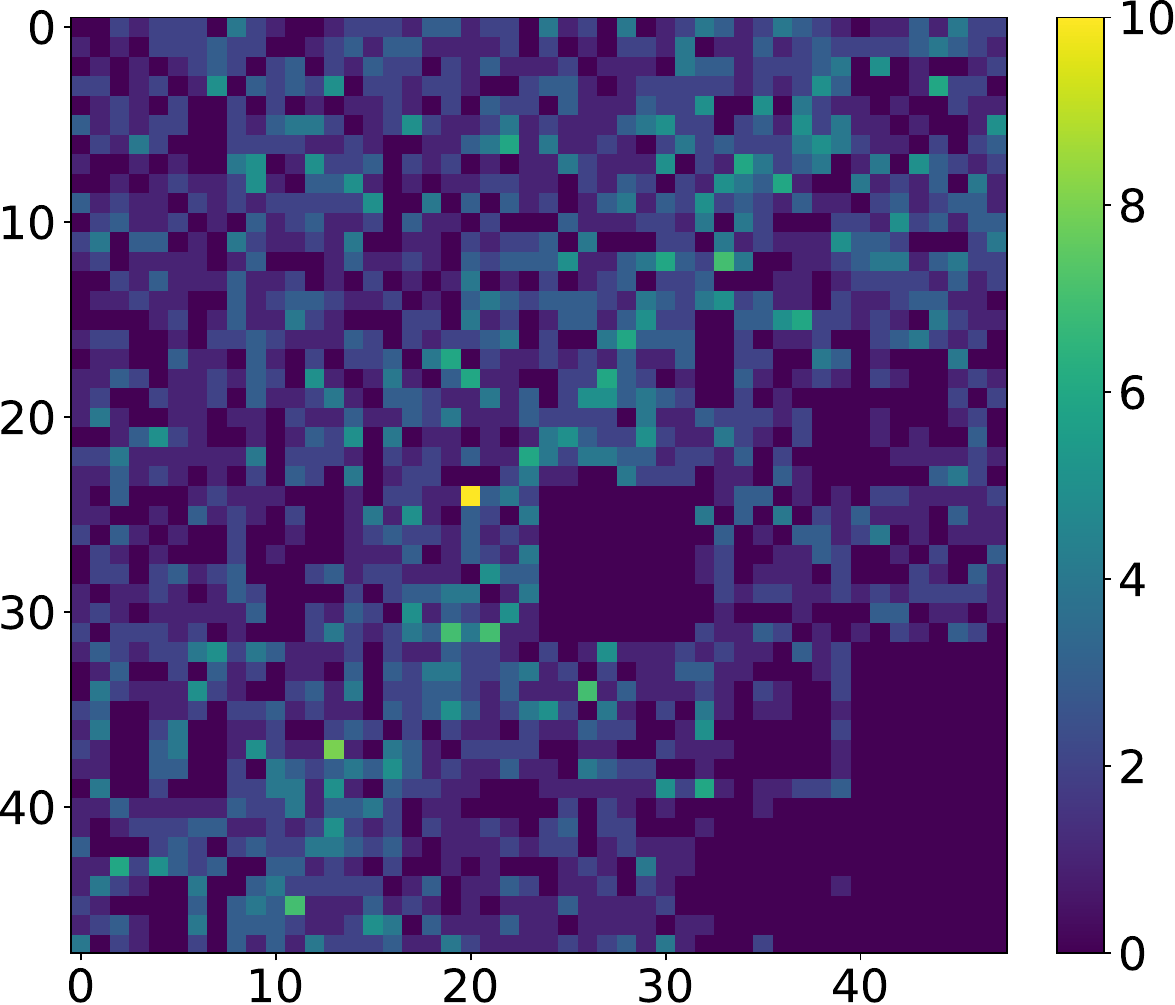}\fig{.245}{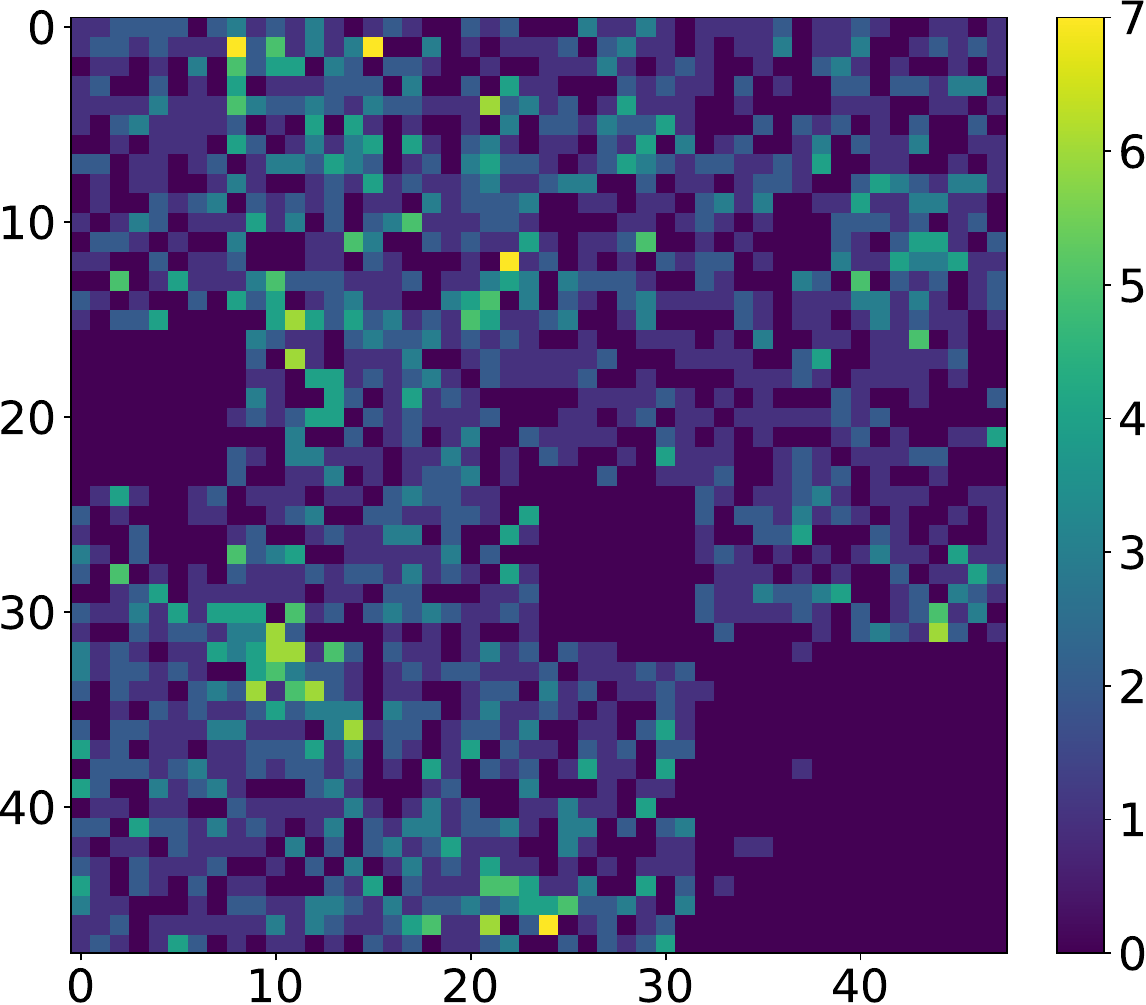}~\fig{.25}{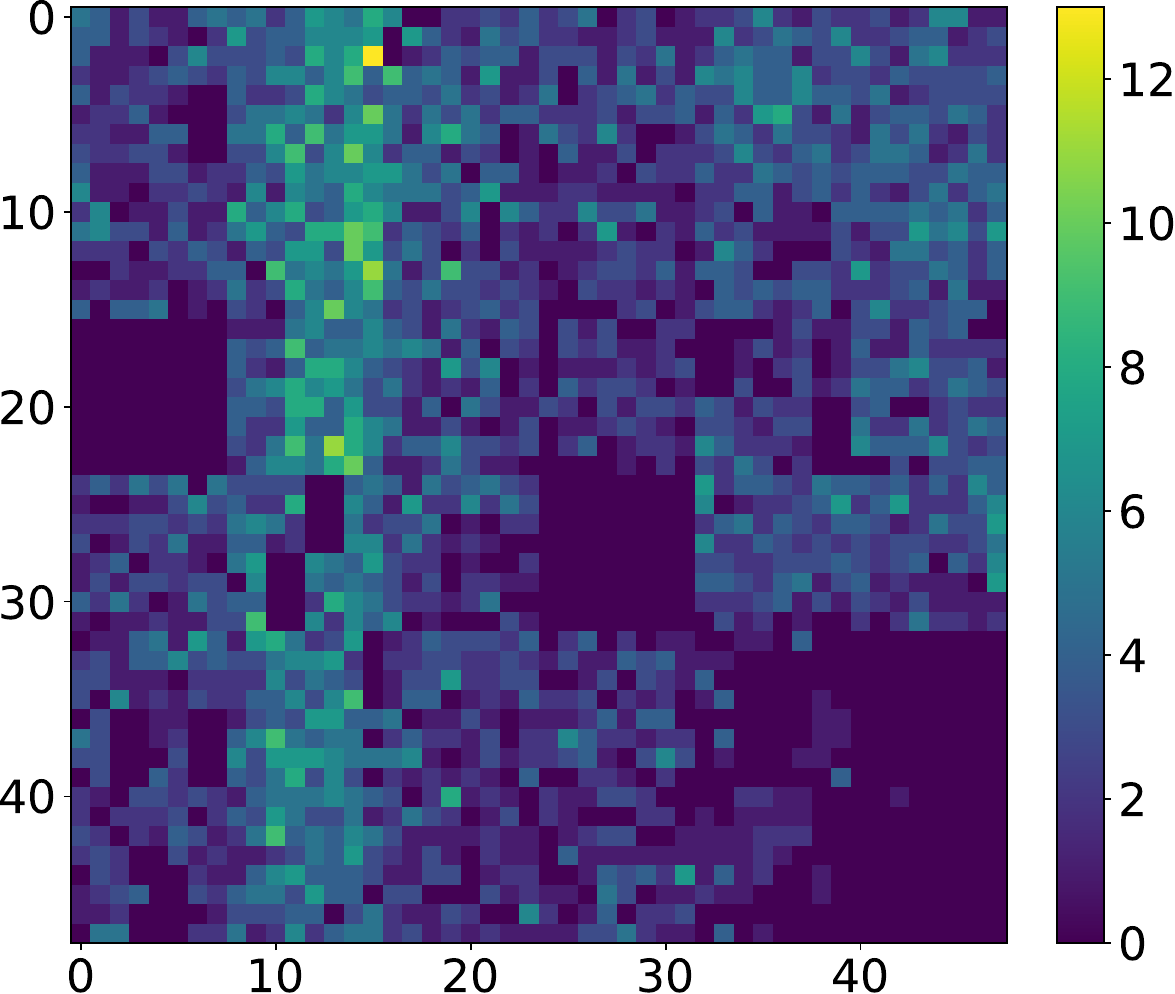}\fig{.254}{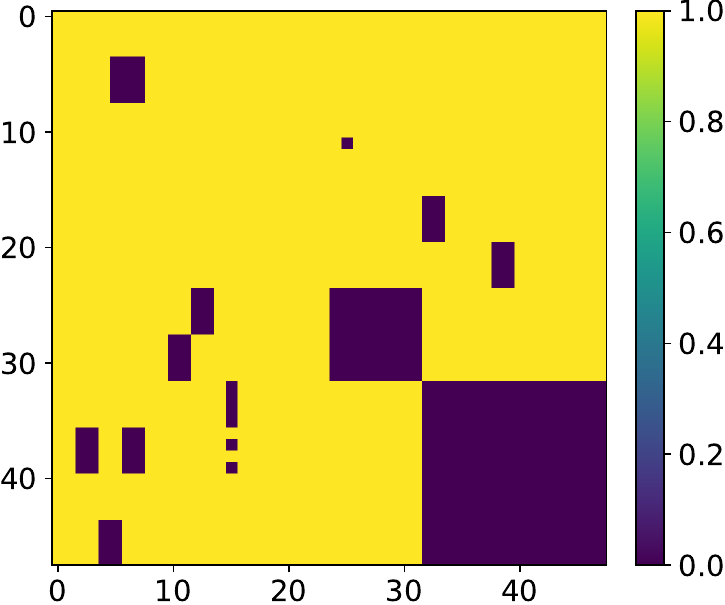}}
	\caption{From left to right: snapshots of the EUSO-TA photodetector
		at the moments of registering UHECRs on 2015-05-13, 2015-10-15,
		2015-11-07, and the mask employed
		to exclude data corresponding to malfunctioning pixels.
		Zeros in the mask indicate pixels that were ignored.}
	\label{events}
\end{figure}

Observations at each elevation angle need training ANNs with dedicated
data sets.  We selected the elevation angle of $15^\circ$ since there
were three UHECRs registered at it. 
Their parameters estimated by the Telescope Array are shown in
Table~\ref{tab:events}~\cite{eusota-2024}. Figure~\ref{events}
presents snapshots of the focal surface of EUSO-TA at the moments
of registering the three events.
The accuracy of energy estimations of the three events is not known.
In what follows, we assume the energy resolution of 17\%, which was
estimated to be typical for monocular observations with the Telescope
Array FTs~\cite{ta-2015}.

\begin{table}[!ht]
	\caption{Parameters of UHECRs registered by EUSO-TA during
	observations with the elevation angle of~$15^\circ$.
	Zenith angles~$\theta$ and azimuth angles~$\phi$ are given in the
	local coordinate system.
	$R_p$ denotes the perpendicular distance from the telescope to the
	EAS axis.}

	\label{tab:events}

	\medskip
	\centering
	\begin{tabular}{|c|c|c|c|c|}
		\hline
		Date & Energy, EeV & $\theta, {}^\circ$ & $\phi, {}^\circ$ &
		$R_p$, km \\
		\hline
		2015-05-13 & 1.15 & 34.5 & 82.8  & 2.5 \\
		\hline
		2015-10-15 & 3.31 & 40.6 & 210.5 & 9.0 \\
		\hline
		2015-11-07 & 2.63 & 8.1  & 8.0   & 2.6 \\
		\hline
	\end{tabular}
\end{table}

\section{Simulated Data}

CONEX~\cite{conex} and EUSO-Offline~\cite{offline} were employed in
Paper~I to prepare data for training and testing the ANNs. We simulated
showers originating from proton primaries using the QGSJET-II-04 model
of hadronic interactions~\cite{qgsjet-2011}. This was justified by some
considerations. First, our models were developed as a proof of concept
aimed to test the possibility of using ANNs for the reconstruction of
UHECR parameters, and thus we did not mean to consider all possible
models of hadronic interactions and mass compositions of UHECRs. On the
other hand, FTs directly observe the longitudinal development of EASs
and have small dependence of event reconstruction on hadronic
interaction models.

In Paper~I, we simulated UHECRs in the energy range 5--100~EeV for
EUSO-TA. However, as it is shown in Table~\ref{events}, events
registered with EUSO-TA had lower energies, and thus the data set was
not appropriate for training an adequate model. Besides this, only data
for the elevation angle of $10^\circ$ were considered. Thus we simulated
a new data set for the elevation angle of $15^\circ$ in the energy range
0.63--6.6~EeV. Energies were distributed uniformly vs.\
$\log(E/\mathrm{eV})$ with the step of 0.02. Due to this, the difference
between any two adjacent energies was $\sim5\%$, which is better than
expected energy resolution. The total data set contained more than 250
thousand events.
Random subsets of different sizes were extracted from it for training
and testing the ANNs, see below.

Simulations were performed in such a way that all shower cores were
located within the projection of the telescope field of view on ground.
The maximum distance from the telescope to EAS cores depended almost
linearly on $\log(E/\mathrm{eV})$ providing approximately the same
trigger efficiency $\sim50\%$ in the whole energy range. However, this
also resulted in non-uniform distribution of events vs.\ energy in the
data set.  Azimuth angles were distributed uniformly from all
directions. Zenith angles varied from $0^\circ$ up to $45^\circ$ with
their number $\sim\cos\theta$, which is the default in CONEX.  The main
quality cut on events included in the data set was the trigger
implemented in EUSO-Offline.

Every event in the EUSO-TA data is represented by a recording that consists
of 128 time frames (gate time units, GTUs) with the time step equal to
2.5~$\mu$s. Each time frame, in turn, contains $48\times48$ integer numbers
specifying photon counts in the pixels of the photodetector.  However,
the maximum number of ``active'' frames containing signals from EAS
tracks in the simulated data did not exceed 11, with less than 0.15\% of
events having more than 8 active frames. Thus, we reduced the records to
just~8 frames. This allowed loading 76 thousand samples in the available
video RAM at the stage of training the CNNs and 86 thousand events
for training the CEDs.\footnote{Since each of the real
events contained just a single active frame, we checked if the quality
of energy reconstruction of such short events can be improved by using
samples consisting of 2 or 4 frames with respective increasing of the
size of the training data set.  This did not happen.}
The latter was possible due to the fact that the CEDs did not need to
consider ``inactive'' frames to recognize activated pixels, so that it
was possible to load only active frames.
All these training sets were selected in a random fashion from the
initial data set, that contained more than 250 thousand events.

The background radiation was simulated to have Gaussian distribution
with the mean equal to 1 photon count per pixel per GTU, which
corresponds to typical observation conditions during clear moonless
nights. We did not take into account possible non-uniformities of the
background illumination due to clouds or stars.

Approximately 17\% of the EUSO-TA pixels did not function properly
during the campaign of 2015, thus we applied a special mask to all
simulated data to make them as close to the real data as possible, see
the rightmost panel in Fig.~\ref{events}.

\section{Reconstruction of Energy of the EUSO-TA Events}
\subsection{Model Training}

Similar to the conventional approach and the method suggested
in~Paper~I, we began with training models for recognition of EAS tracks.
The task can be considered as semantic segmentation: each pixel in each
frame of a sample must be assigned to a specific class.  In our case,
there are only two classes (types) of pixels: those that belong to an
EAS track (activated pixels) and all the rest. We employed the
convolutional encoder-decoder used in Paper~I to accomplish the task.
To choose the best model, we trained 10 models with different initial
weights. Each model was trained on its own data set consisting of 86
thousand events randomly chosen from the initial set.
Before processing, each event was scaled to
$[0, 1]$, and pixels with values above~0.28 were marked as hit pixels.
This was justified by an analysis of the dependence of performance of
models on the threshold.
The models were tested on the same data set consisting of 1000 events.

The quality of the models was compared using three performance metrics
that are widely used in such tasks: PR AUC (area under the
precision-recall curve), balanced accuracy, and mean
intersection-over-union.  The best model had the following values of
these metrics: 0.843, 0.848, and 0.788 respectively. (All three
functions used to change in a similar fashion, thus it was sufficient to
use any one of them.) The selected model was used next to mark tracks in
the whole data set.\footnote{Our tests have demonstrated that one can
use other data sets to train CEDs. This means that one can use, say,
set~A to train a CED, and then employ the CED to recognize activated
pixels in data set~B, which will be used then to train a CNN for energy
reconstruction.}
Remarkably, models trained this way could also be used for searching for
EAS tracks in data of the PAIPS-L fluorescence
telescope~\cite{paips-2025}.

After labeling activated pixels in the simulated data set, we trained the
6-layer CNN introduced in Paper~I to reconstruct energy of primary
UHECRs.  Three ways of training models for energy reconstruction were
tested.  First, we trained models that had only energy as a parameter to
be predicted (``Models E'' below). Next, we added distance from the
telescope to the shower core as an additional parameter (``Models (E,
dist)'').  Finally, we added zenith and azimuth angles (``Models (E,
dist, $\theta$, $\phi$)''). All models were trained using the same
labels of hit pixels obtained with the selected CED model.
The quality of energy reconstruction was estimated based on the mean
absolute percentage error (MAPE) calculated for test samples consisting
of 1000 events selected in a random fashion from the initial data set.
Coefficient of determination $R^2$ was used as an additional performance
metric.

\begin{figure}[!ht]
	\centerline{\fig{.493}{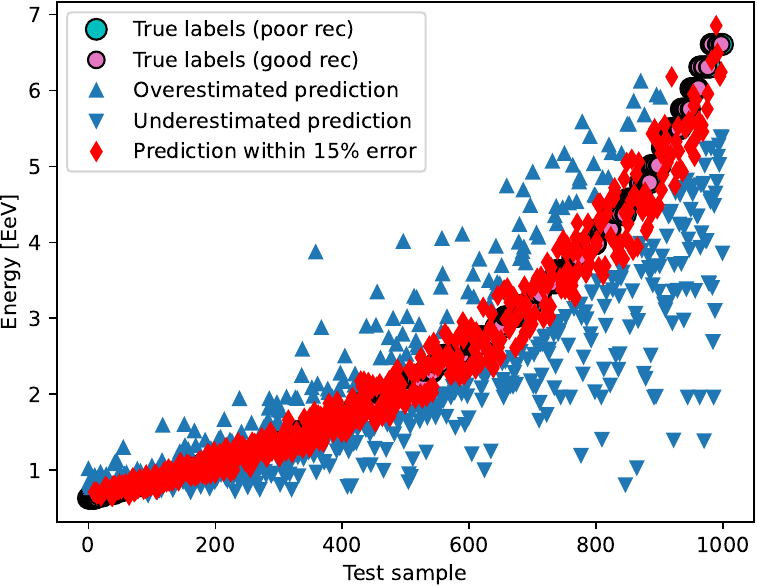}~\fig{.49}{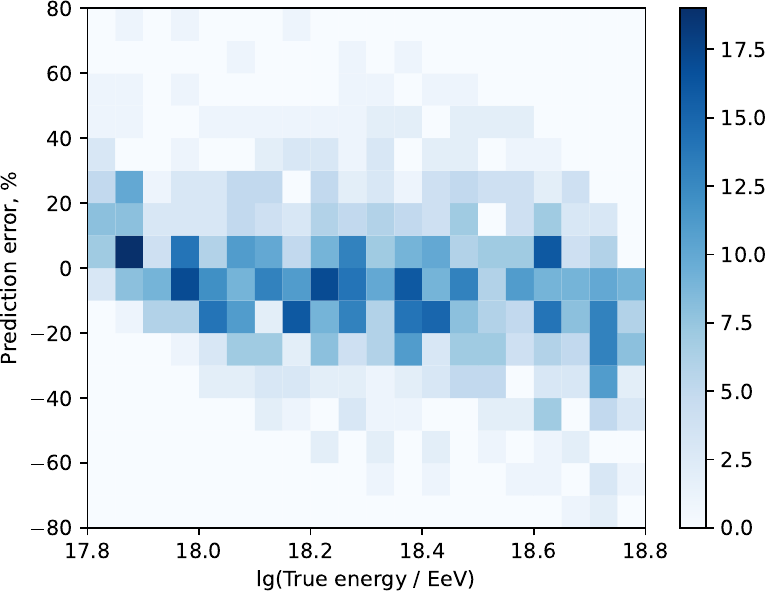}}

	\caption{An example of energy reconstruction for 1000 events with a CNN
	trained to predict only energy (one of ``Models~E''). Left: predictions of
	the model. Ground truth labels are indicated by circles. Red diamonds show
	predictions that are within 15\% error from the true ones. Blue triangles
	pointed up and down show predictions overestimated and underestimated by
	more than 15\% respectively. Right: distribution of relative errors over
	energy of events in the test sample.}

	\label{ysems}
\end{figure}

Figure~\ref{ysems} presents an example of energy reconstruction
for a sample consisting of 1000 events. The model was trained to predict
only energy (i.e., this was one of ``Models~E''). Active pixels were
marked with the help of the convolutional encoder-decoder described
above. Mean absolute percentage error for these particular model and
test set was equal to 17.8\% with the bias and the standard deviation
equal to $-4.0\%$ and 23.3\% respectively. It is clearly seen from both
panels of Figure~\ref{ysems} that predicted values of energy are
underestimated in average, while at the lower end of energies in the
test sample predictions tend to be overestimated. This kind of behavior
was observed in all tests. It was also observed in Paper~I for the
data set consisting of events in the energy range 5--100~EeV.

\begin{figure}[!ht]
	\centerline{\fig{.493}{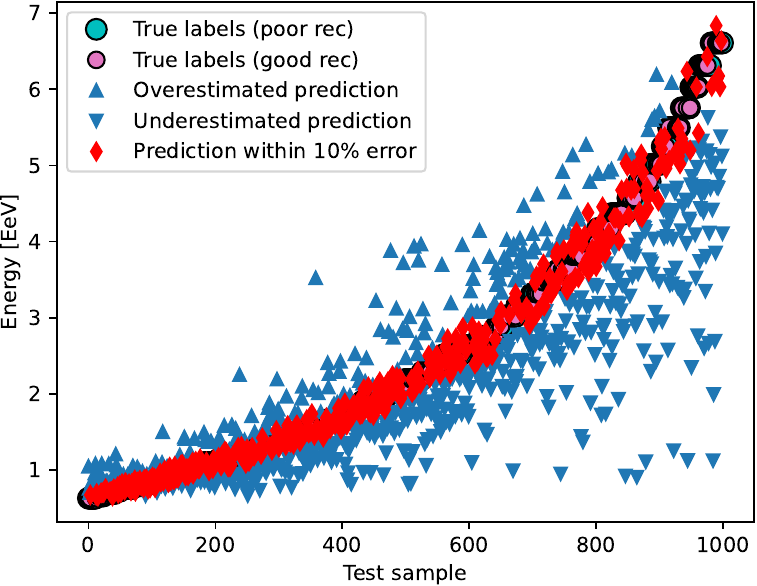}~\fig{.49}{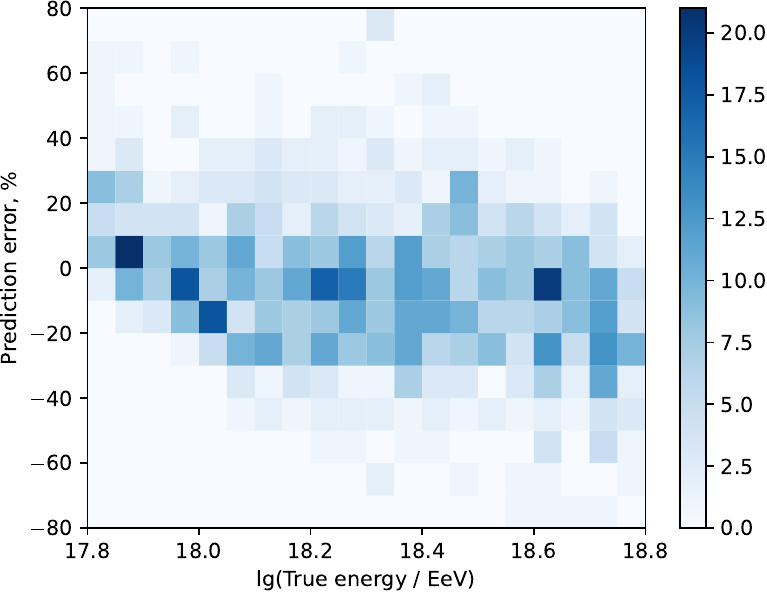}}

	\caption{An example of energy reconstruction for 1000 events
	with a CNN trained to predict energy without any
	information about EAS tracks. Notation is
	the same as in Fig.~\ref{ysems}.
	}

	\label{blind}
\end{figure}

Once we made a mistake and passed unlabeled data to the CNN.  In this
case the CNN aimed to reconstruct energy had no information about
tracks but just pure data. Surprisingly, the performance of the model
was not much worse than that of models trained in the standard
way.  Figure~\ref{blind} presents results of applying a CNN trained in a
``blind'' way to the same test set as shown in Fig.~\ref{ysems}.  In
this case, MAPE was equal to 18.4\% with the bias and the standard
deviation equal to $-5.3\%$ and 23.7\% respectively.

We faced one important problem with all models that we trained:
predictions of energy for particular events could noticeably vary even
for models with practically the same MAPE, $R^2$, and biases.  We tested
several other performance metrics but failed to find any one that would
allow us to select the most accurate model.  Seemingly the easiest way
to overcome this problem is to train a sufficiently large number of
models, obtain estimates of energy of any given event with each of them,
and then take the respective mean values. The standard deviation can be
used then as a measure of accuracy of the estimate (though certain
restrictions of this approach must be kept in mind).
The estimates can also be corrected taking into account the mean bias.

\begin{figure}[!ht]
	\centerline{\fig{.5}{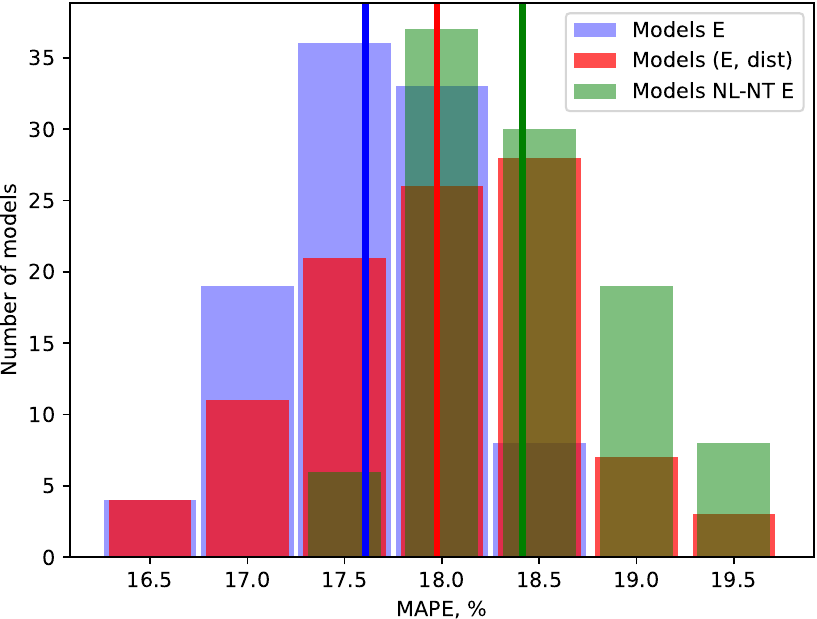}}

	\caption{Distribution of MAPE in predictions of energy for three sets
	of models: models trained to reconstruct only energy with tracks
	recognized by the CED (Models E), models trained to reconstruct both
	energy and the distance from the FT to the shower core with tracks
	recognized the same way (Models (E, dist)), and models trained to
	reconstruct energy without any knowledge about EAS tracks (NL-NT~E).
	Vertical lines of three colors indicate mean values of MAPE for the
	respective sets.}

	\label{mape}
\end{figure}

Figure~\ref{mape} shows the distribution of mean absolute percentage
errors of energy estimates calculated for models trained to reconstruct
only energy (Models~E), to reconstruct both energy and the distance from
the telescope to the shower core (Models (E, dist)), and models trained
to reconstruct energy without any information about EAS tracks (Models
NL-NT E, ``No Labels--No Tracks''). One hundred models were trained in
each case. The mean values of MAPE are equal to 17.6\%, 18.0\%, 18.4\%
respectively with the standard deviations of MAPEs equal to 0.5\%,
0.7\%, and 0.5\%. In all cases the mean bias of estimates was around
$-3.4\%$ with the median values around $-4.5\%$.

\subsection{Event registered on 2015-05-13}

\begin{figure}[!ht]
	\fig{1}{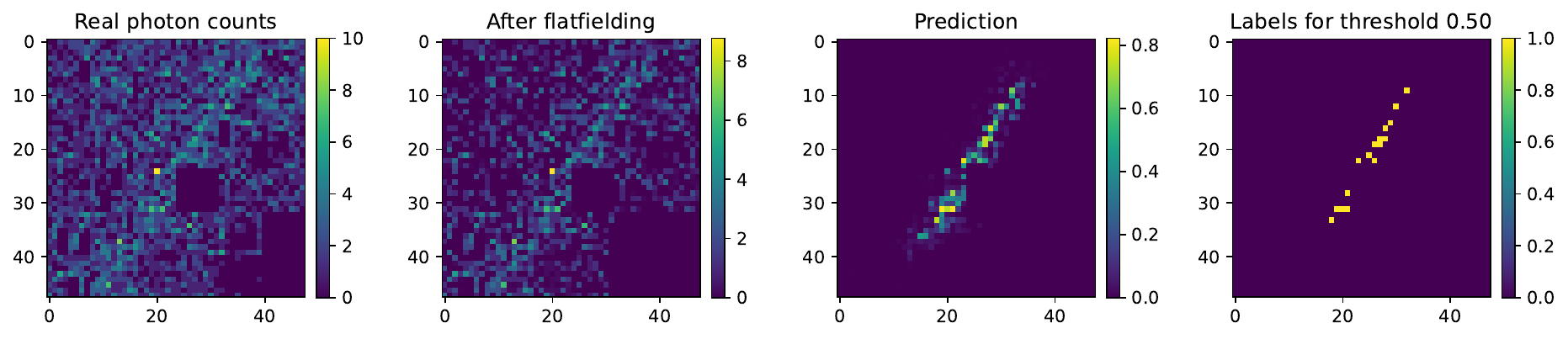}

	\caption{Recognition of active pixels in the event registered on
	2015-05-13.  From left to right: the original signal; the
	flat-fielded signal (with the mask applied); prediction of the CED;
	activated pixels as predicted with the threshold level 0.5.}

	\label{20150513}
\end{figure}

Figure~\ref{20150513} illustrates the process of recognizing the track
of the event registered on 2015-05-13. The left panel presents the
original signal over the photodetector (the same as in
Fig.~\ref{events}). The next panel shows results of flat-fielding of the
original signal. To do this, mean values of photon counts were
calculated for each pixel during the previous 256 GTUs and then
extracted from the actual values.  Negative values were zeroed. The mask
shown in the leftmost panel of Fig.~\ref{events} was also applied. The
third panel shows probabilities of pixels to belong to the track as
found by the CED. Finally, the fourth panel shows the EAS track as
recognized by the model. One can see that the track is not continuous,
which can be the result of a shortcoming of the model as well as the
existence of several malfunctioning pixels.

\begin{figure}[!ht]
	\centerline{\fig{.49}{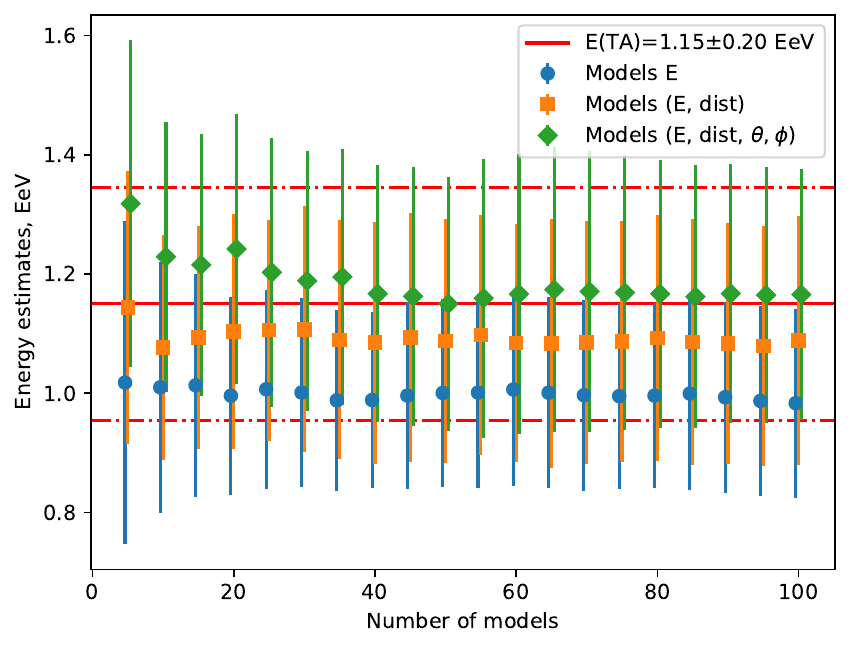}\quad\fig{.49}{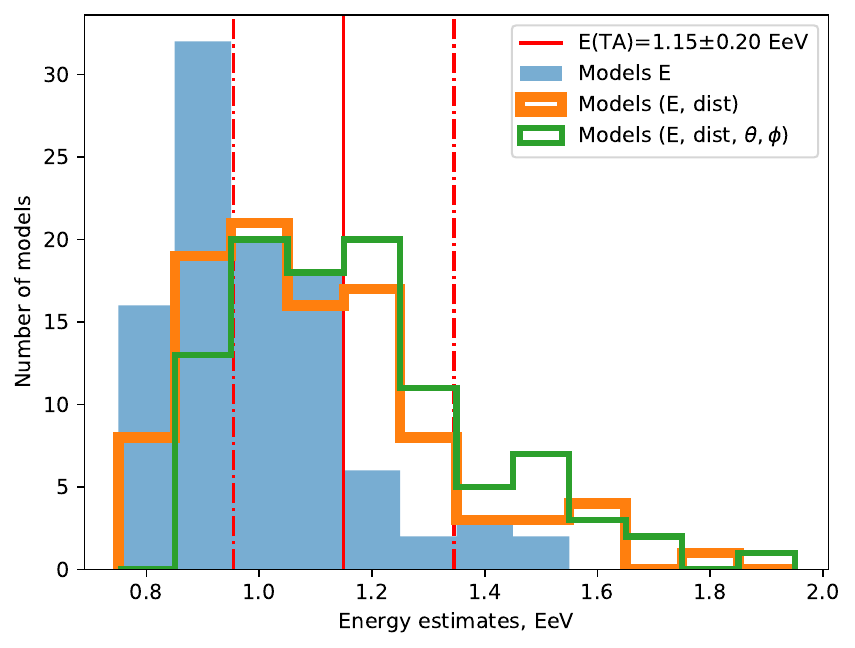}}

	\caption{Mean estimates of energy (left) and distribution of
	estimates (right) for the UHECR registered on 2015-05-13 obtained with
	three different sets of models. Data points in the left panel are
	slightly shifted vs.\ each other to improve visibility.  See the text
	for other details.}

	\label{20150513est}
\end{figure}

The left panel of Fig.~\ref{20150513est} presents results of obtaining
energy estimates of the event 2015-05-13 with 100 models of each of the
three types described above that employ information about activated
pixels during training. Data points indicate mean estimates obtained
with the respective number of models. Errors are calculated as standard
deviations\footnote{A more complicated approach to estimating errors
might be needed due to the non-Gaussian distribution of model
predictions.}.
One can see how mean estimates converge to a certain value
after applying approximately 60 models and almost stop varying as
training of more models continues. A similar behavior is observed if the
order of models is mixed in a random fashion. The right panel of
Fig.~\ref{20150513est} shows the distribution of estimates of energy of
the UHECR in the three sets of models.

Table~\ref{tab20150513} provides numerical values of energy estimates,
as well as median values and estimates obtained after excluding outliers
with the Tukey fence\footnote{In this case, we only keep estimates
within $[Q_1 - k(Q_3-Q_1); Q_1 + k(Q_3-Q_1)]$, where $k=1.5$, and $Q_1$
and $Q_3$ are the lower and upper quartiles respectively.} (``fixed''
mean and median).  In each case, the models used the track shown in
Fig.~\ref{20150513}.  It can be easily seen that estimates obtained with
all three sets of models are reasonably good being within $\pm0.20$~EeV
energy resolution of monocular observations with FTs at the Telescope
Array.  The accuracy of predictions estimated as the standard deviation
is also of the same order.  The best estimates are obtained as the mean
values of predictions of models trained with four parameters: (E, dist,
$\theta$, $\phi$). 

\begin{table}[!ht]
	\caption{Estimates of energy of the UHECR registered on 2015-05-13.
	The Telescope Array collaboration estimated its energy as
	$1.15\pm0.20$~EeV. All values are given in units of~EeV.
	See the text for details.}

	\label{tab20150513}

	\medskip
	\centering
	\begin{tabular}{|l|c|c|c|c|c|}
		\hline
		Models & E & (E, dist) & (E, dist, $\theta$, $\phi$) & NL-NT~E &
		NL-T~E \\
		\hline
		Pure mean$\pm$std & $0.98\pm0.16$ & $1.09\pm0.21$ & $1.17\pm0.21$
		                  & $0.78\pm0.05$ & $1.10\pm0.53$ \\
		Pure median       & 0.95 & 1.06 & 1.13 & 0.78 & 1.02 \\
		Fixed mean$\pm$std
						      & $0.95\pm0.11$ & $1.03\pm0.14$ & $1.10\pm0.14$
						      & $0.77\pm0.04$ & $0.96\pm0.40$ \\
		Fixed median      & 0.93 & 1.02 & 1.07 & 0.78 & 0.92 \\
		\hline
	\end{tabular}
\end{table}

Table~\ref{tab20150513} contains two other columns that present
estimates based on 100 models trained to reconstruct energy in a
``blind'' manner, without any information about EAS tracks.  Estimates
shown in the NL-NT~E (No Labels--No Tracks) column were obtained by just
passing the experimental data to these models. Estimates shown in the
NL-T~E (No Labels--Track) column were derived by recognizing the track
in the experimental data first and then applying models for energy
reconstruction. One can see that the first method underestimated the
energy, but estimations were very ``stable'' with only minor variations.
The second method provided a much better estimation on par with the (E,
dist) models but the variance was pretty large.  Anyway, both estimates
look quite acceptable.  In all cases, excluding outliers with the Tukey
fence did not help improving the estimates but just reduced their
variance.  Adding the distance to the shower core and arrival directions
to the number of parameters used to train NL models did not noticeably
affect the results.

\subsection{Event registered on 2015-11-07}

\begin{figure}[!ht]
	\fig{1}{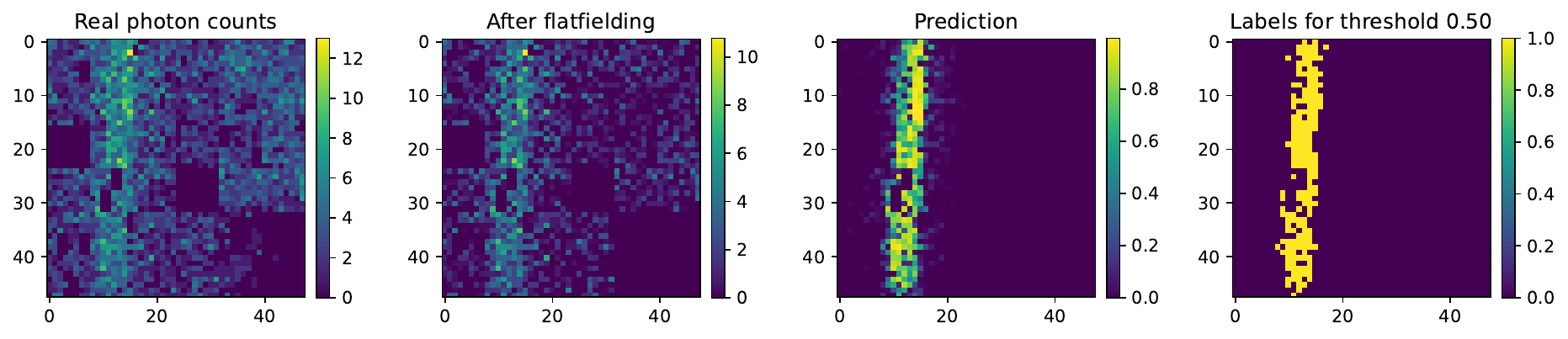}

	\caption{Recognition of activated pixels in the event registered on
	2015-11-07.  From left to right: the original signal; the
	flat-fielded signal (with the mask applied); prediction of the CED;
	activated pixels as predicted with the threshold level 0.5.}

	\label{20151107}
\end{figure}

Figure~\ref{20151107} demonstrates the process of recognizing activated
pixels for the UHECR registered on 2015-11-07. Similar to
Fig.~\ref{20150513}, one can see the original signal, its flat-fielded
version passed to the CED, prediction of the CED, and pixels labeled as
the track.

\begin{figure}[!ht]
	\centerline{\fig{.49}{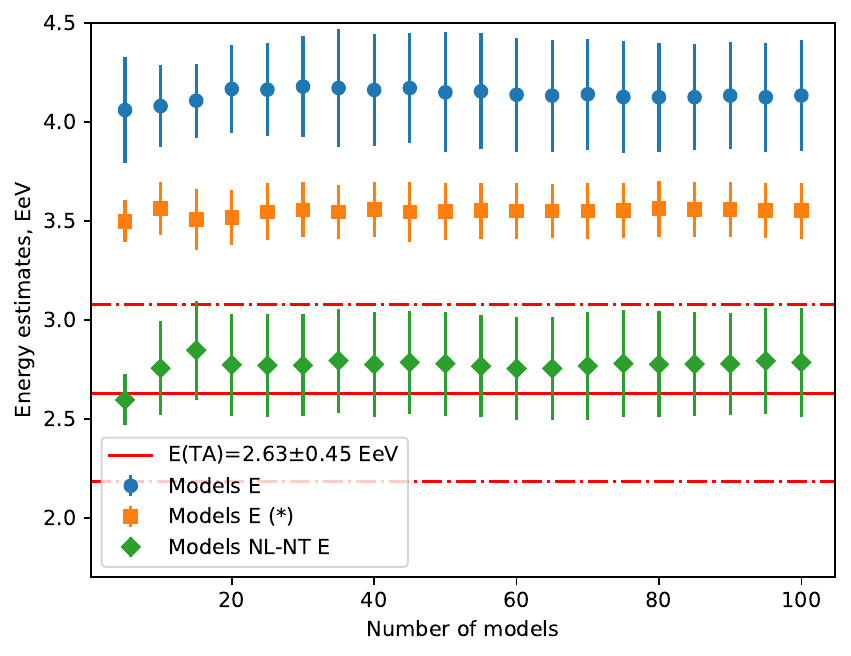}\quad\fig{.49}{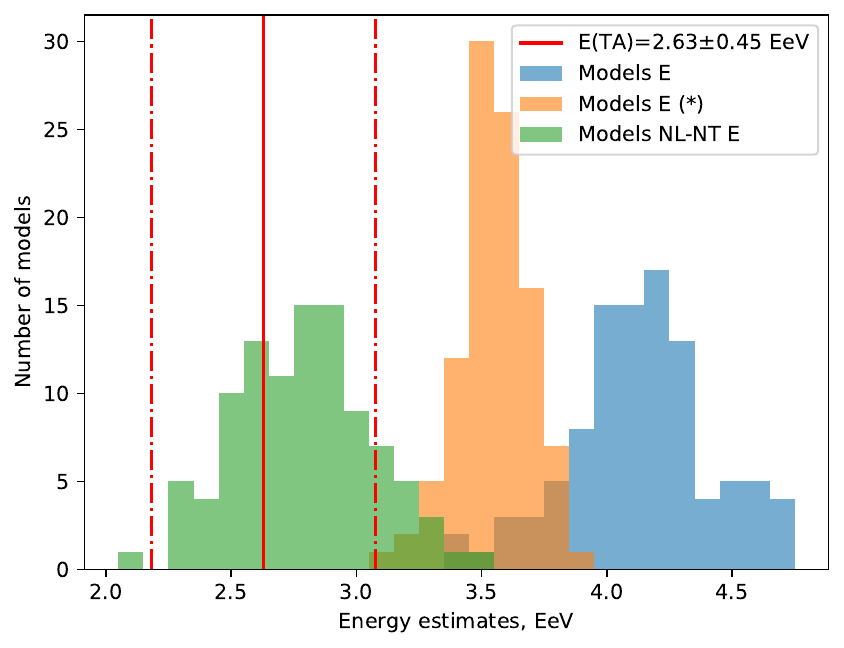}}
	\caption{Mean estimates of energy (left) and distribution of estimates
	(right) of the UHECR registered on
	2015-11-07 obtained with three different sets of models.
	See the text for details.}
	\label{20151107est}
\end{figure}

Despite the clearly pronounced track, the event posed a problem for
reconstruction of its energy. All three types of models that were
trained with information about activated pixels, provided estimates
in the range $\approx4.1$--4.3~EeV,
much higher than 2.63~EeV by the Telescope Array, see
an example for Models~E in Fig.~\ref{20151107est}.  Surprisingly, models
NL-NT~E trained and applied without any knowledge about tracks, resulted
in a much better estimate $2.79\pm0.27$~EeV ($2.73\pm0.23$~EeV with
outliers excluded), see Table~\ref{tab20151107}.

\begin{table}[!ht]
	\caption{Estimates of energy of the UHECR registered on 2015-11-07.
	The Telescope Array collaboration estimated its energy as
	$2.63\pm0.45$~EeV. All values are given in units of~EeV.
	See the text for details.}

	\label{tab20151107}

	\medskip
	\centering
	\begin{tabular}{|l|c|c|c|c|}
		\hline
		Models & E & E(*) & NL-NT E & NL-NT E(*) \\
		\hline
		Pure mean$\pm$std & $4.13\pm0.28$ & $3.55\pm0.14$ & $2.79\pm0.27$
		                  & $2.77\pm0.21$ \\
		Pure median       & 4.14 & 3.55 & 2.78 & 2.75 \\
		Fixed mean$\pm$std
						      & $4.08\pm0.20$ & $3.54\pm0.11$ & $2.73\pm0.23$
						      & $2.70\pm0.15$ \\
		Fixed median      & 4.09 & 3.53 & 2.76 & 2.71 \\
		\hline
	\end{tabular}
\end{table}

The discrepancy between the two estimates made us take a closer look at
the training data set. We have found that our data set did not contain a
single nearly-vertical event with the shower core close to the
telescope. All signals with similar brightness were produced by EASs
originating from protons with much higher energies but with the shower
cores located at distances $>3$~km from the FT. To address the issue, we
simulated more than 3500 nearly vertical events with the cores located
within 3~km from the aperture. To increase their ``visibility'' in the
training data set, we took a subset of the initial data set and added
the newly simulated events to it. The new data set contained 100
thousand events. Results obtained with it are indicated with (*).
Besides the results discussed above, Table~\ref{tab20151107} shows
estimates obtained with Models~E(*) and NL-NT E(*), i.e., similar models
but trained on the reduced data set complemented with nearly vertical
showers. It can be seen that while ``blind'' estimates remained almost
intact, values obtained with recognition of tracks became closer to the
true value but still overestimated it.

\subsection{Event registered on 2015-10-15}

\begin{figure}[!ht]
	\centerline{\fig{1}{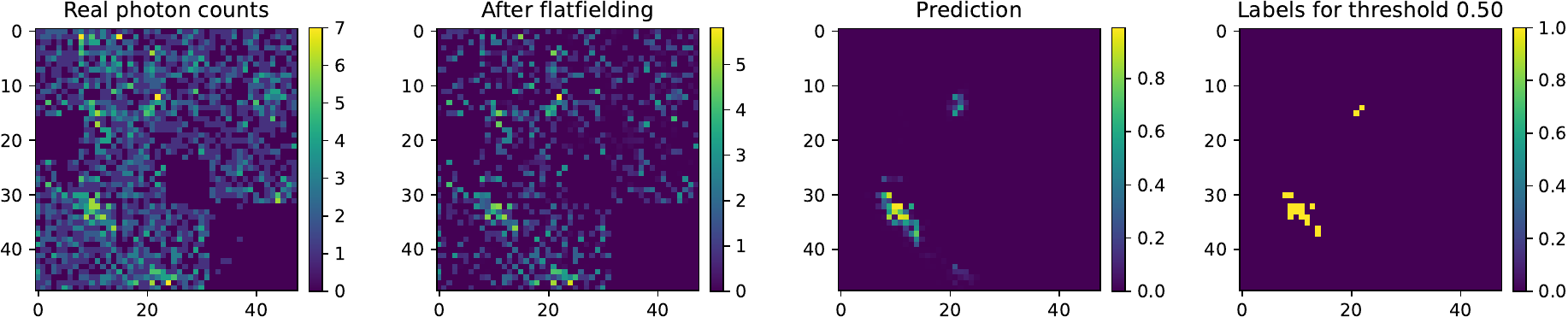}}
	\centerline{\fig{.225}{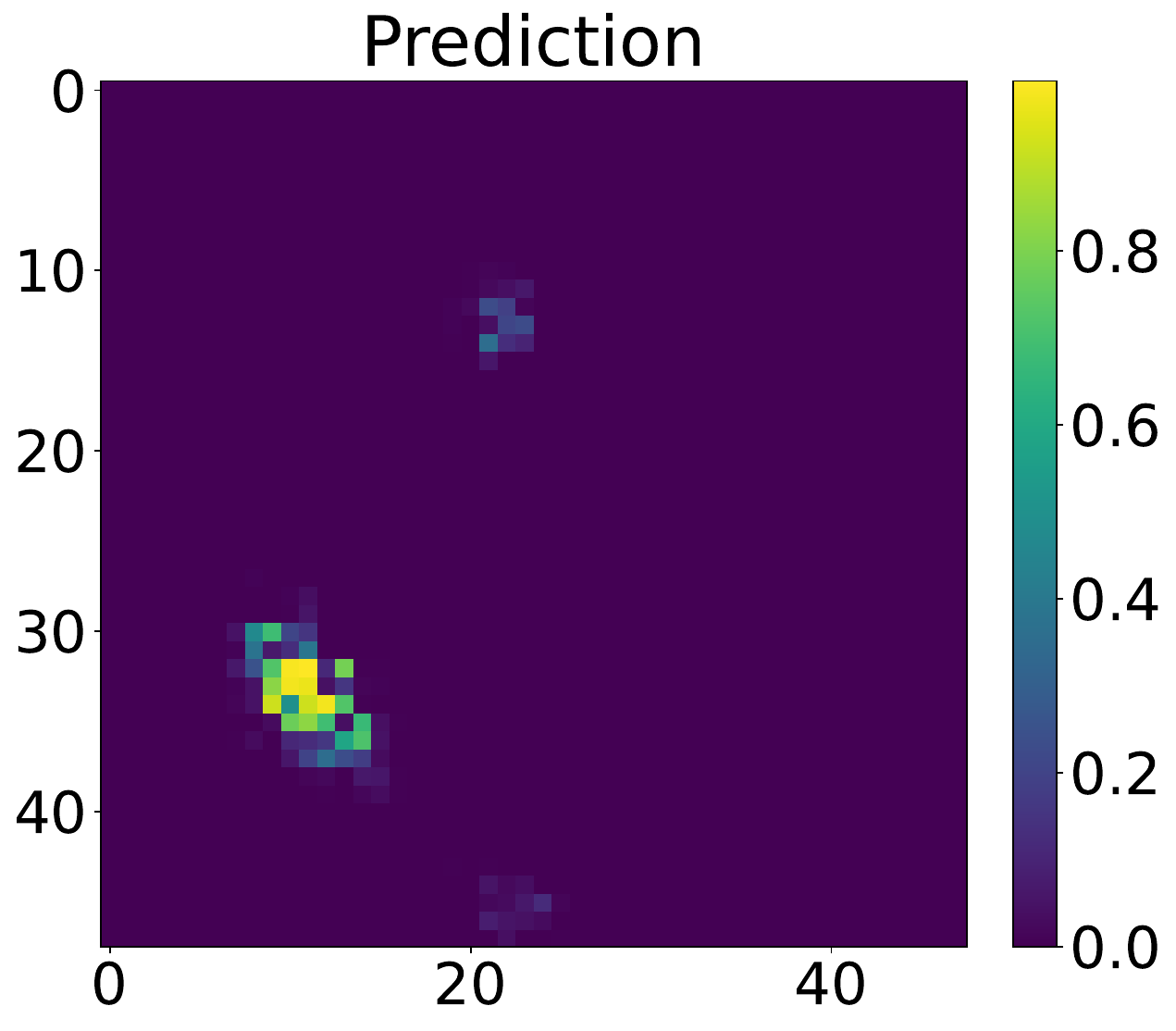}\quad\fig{.225}{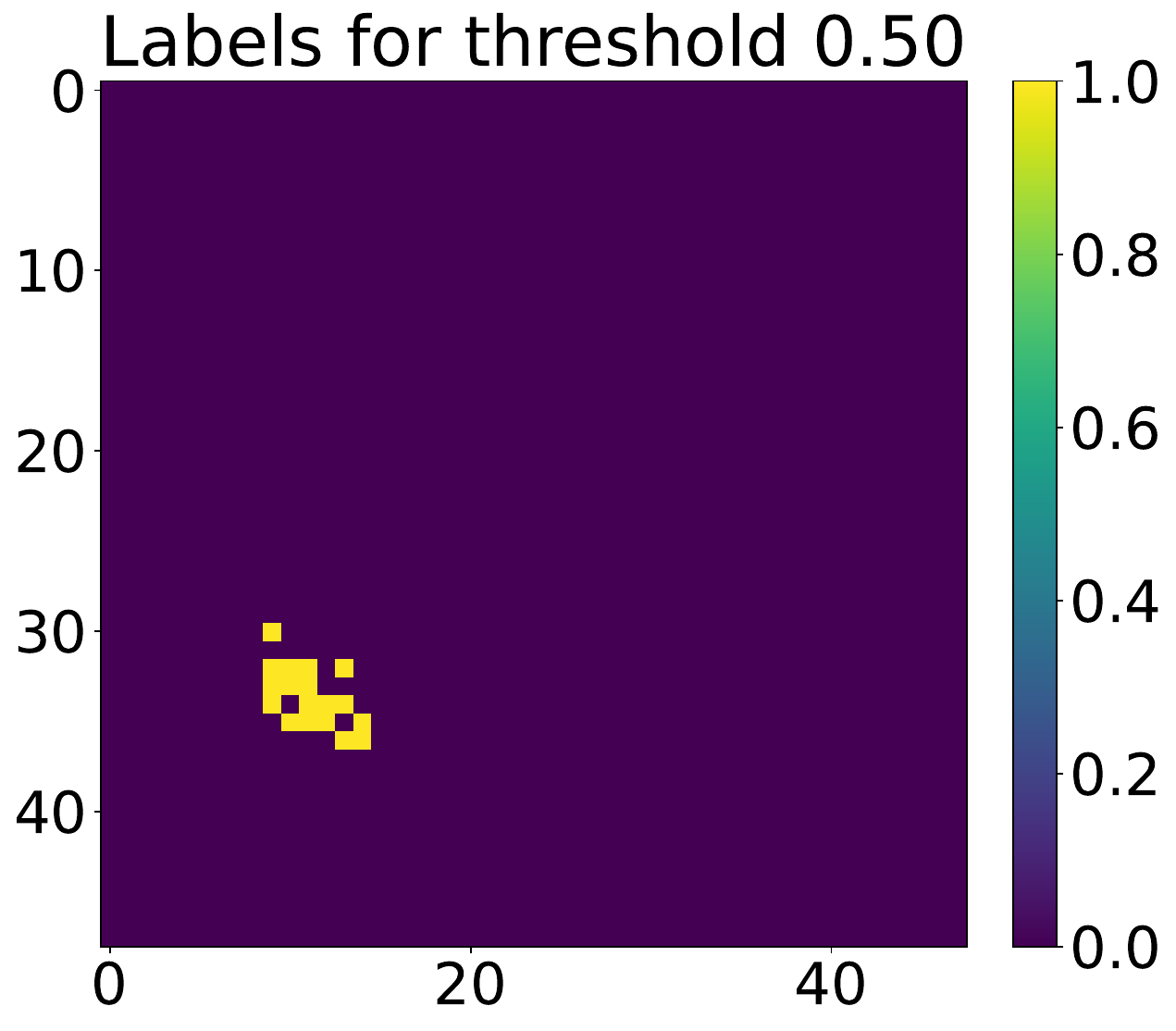}\quad\fig{.225}{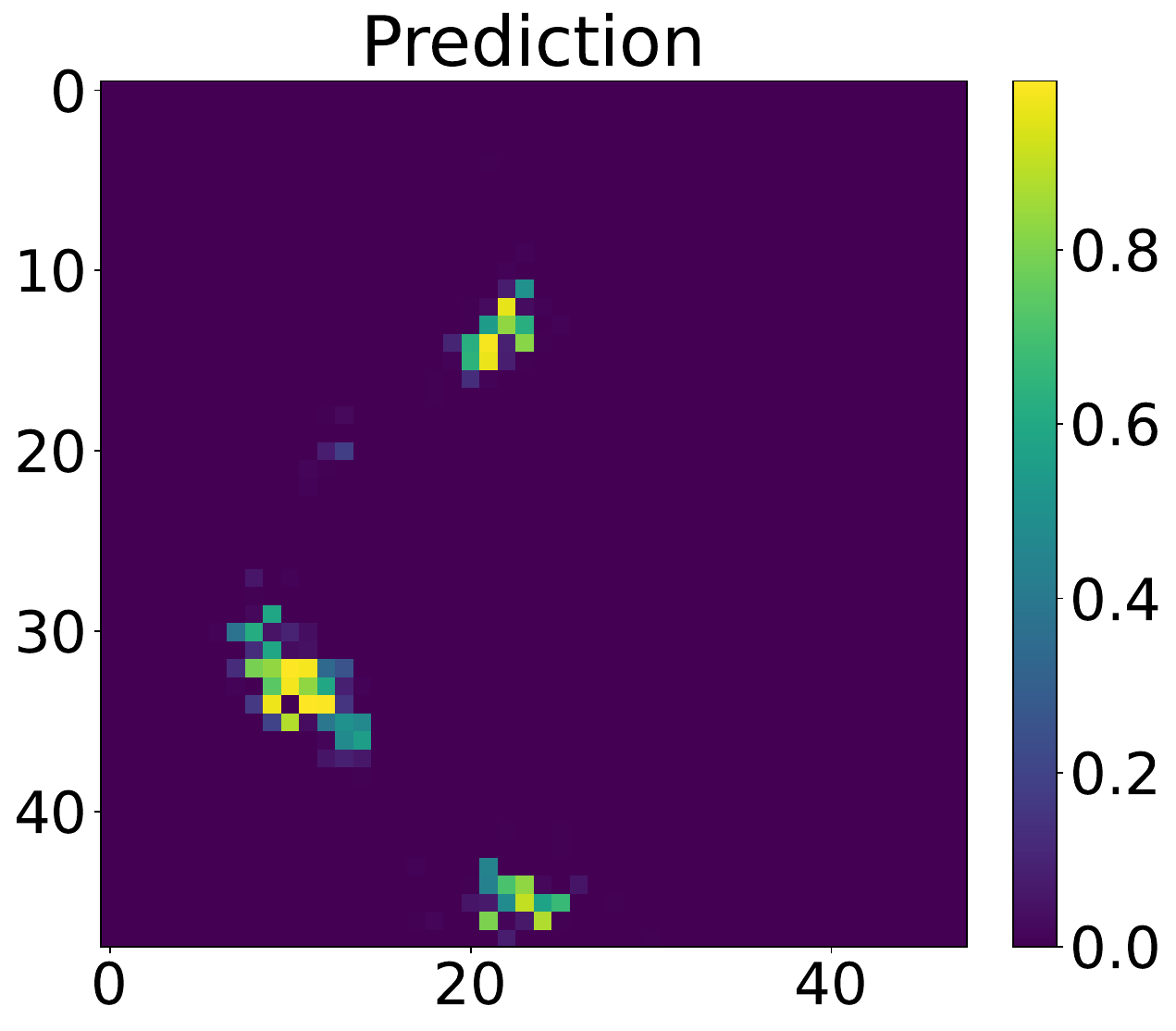}\quad\fig{.225}{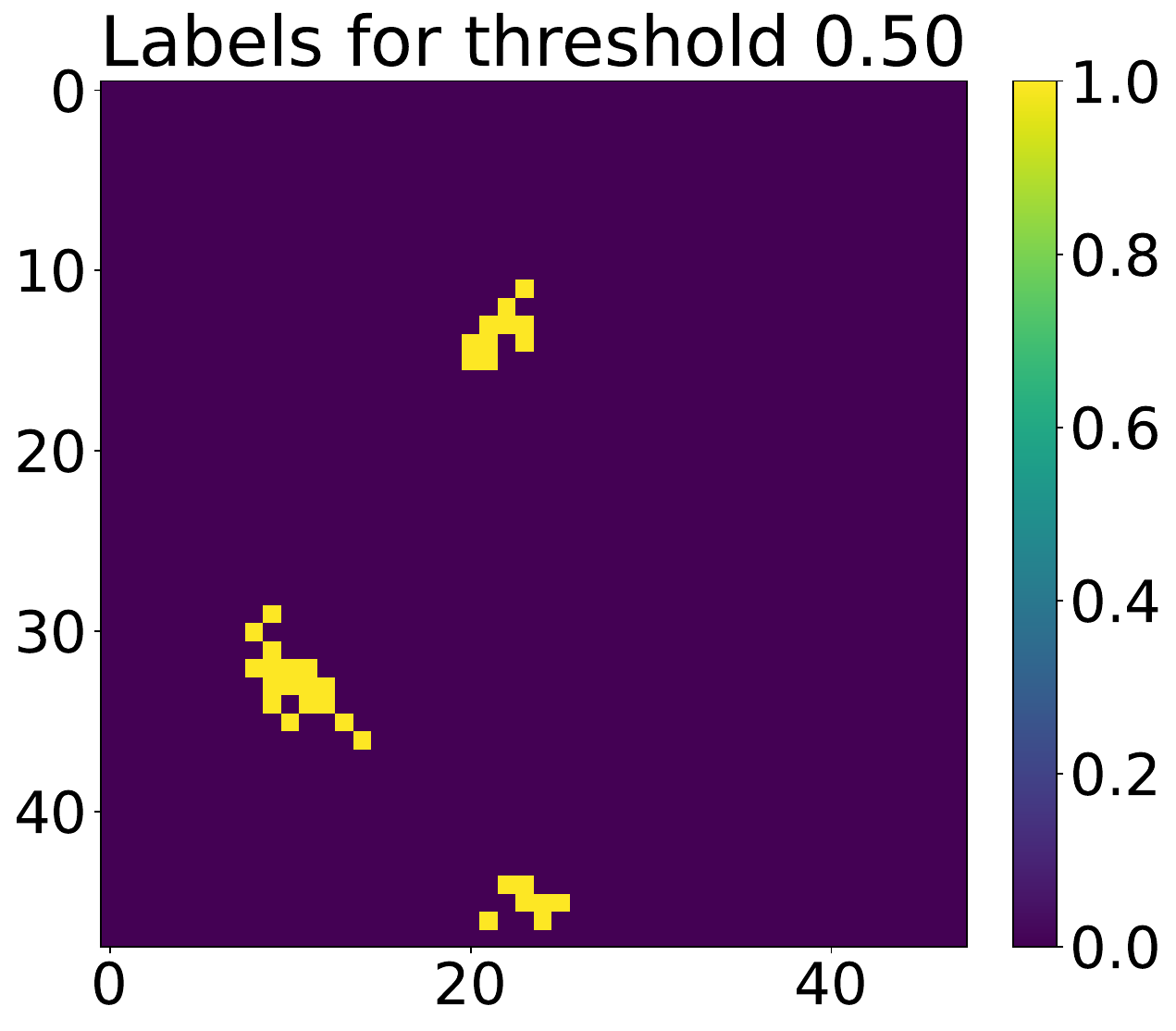}}

	\caption{Recognition of the EAS track of the UHECR registered on
	2015-10-15. The top row, from left to right: the original signal; the
	flat-fielded signal (with the mask applied); prediction of the CED;
	hit pixels as predicted with the threshold level 0.5.
	Bottom row: the same track as recognized by two other models. The
	first and the third panels show predictions of the models; the
	second and the fourth panels show the respective selected pixels.
	}

	\label{20151015}
\end{figure}

Recognition of the EAS track in the event registered on 2015-10-15 posed
a certain problem.  The process is demonstrated in Fig.~\ref{20151015}.
The model used above recognized a small group of pixels as a track but
also marked two pixels that could not be a part of the track, see pixels
around $(i,j)=(14,21)$ in the rightmost panel of the top row of
Fig.~\ref{20151015}. We tried a number of other models and found that
while the artifact does not necessarily exist, as shown in the two left
panels of the second row of Fig.~\ref{20151015}, the track might be
longer but it is never recognized as a whole but consists of two parts,
see the two right panels of Fig.~\ref{20151015}.  This suggests that our
models of track recognition do not select hit pixels properly in this
event, so that a part of them is probably missing.  Thus it did not come
as a big surprise when we found that all our models of energy
reconstruction strongly underestimate the energy of the UHECR.

\begin{figure}[!ht]
	\centerline{\fig{.49}{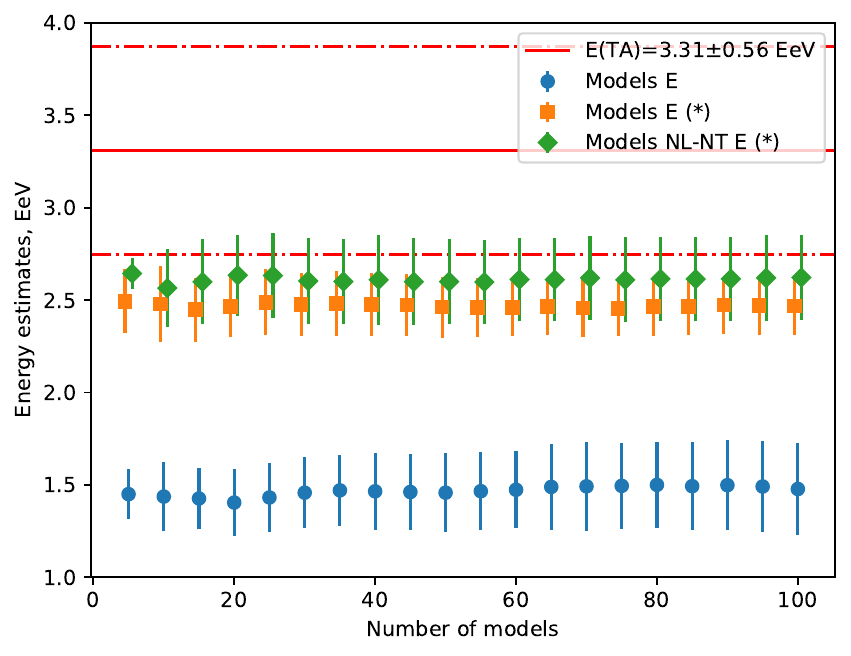}\quad\fig{.49}{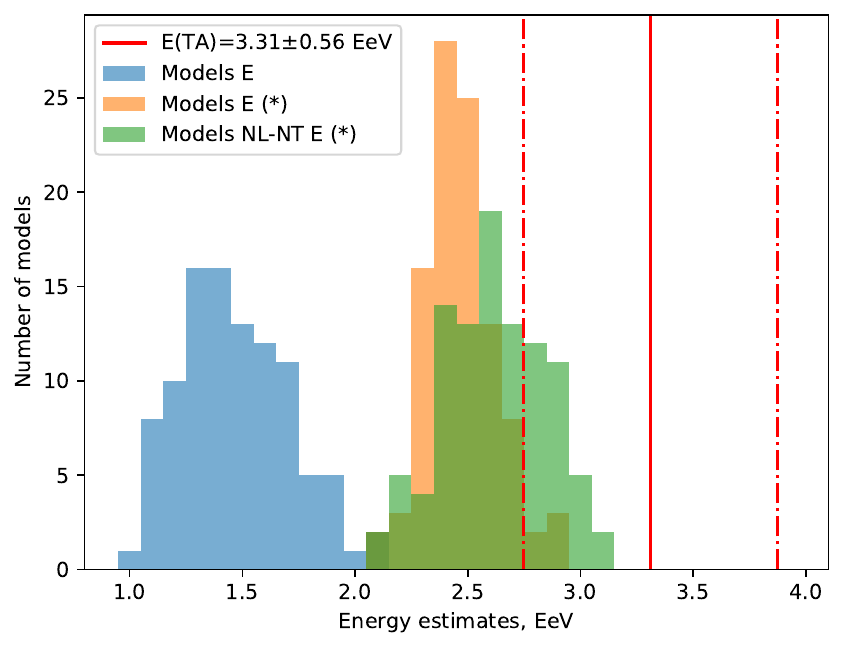}}

	\caption{Mean estimates of energy (left) and distribution of energy
	estimates (right) of the UHECR registered on 2015-10-15 obtained with
	three different sets of models. Data points in the left panel are
	slightly shifted vs.\ each other to improve visibility.  See the text
	for other details.}

	\label{20151015est}
\end{figure}

Models trained with the main training data set to reconstruct only
energy, underestimated the energy of the UHECR by approximately 55\%
giving mean values around 1.48~EeV. Models trained to reconstruct both
energy and the distance to the shower core provided slightly better
estimates with the mean equal to 1.76~EeV. The best estimates were
obtained with the (E, dist, $\theta$, $\phi$) models.
Their mean estimates were equal to 2.09~EeV, i.e., 37\%
below the true value, see Fig.~\ref{20151015est} and
Table~\ref{tab20151015}.  Surprisingly, better estimates were obtained
with models E(*) trained to reconstruct energy on the reduced data set,
and even better ones were provided by models NL-NT E(*) trained in a
blind fashion on the same reduced data set. In the latter case, the mean
estimates of energy were equal to 2.62~EeV, i.e., 21\% below the true
level, see Table~\ref{tab20151015}.
Still, none of the models gave estimates within accuracy of the value
provided by the Telescope Array.

\begin{table}[!ht]
	\caption{Estimates of energy of the UHECR registered on 2015-10-15.
	The Telescope Array collaboration estimated its energy as
	$3.31\pm0.56$~EeV. All values are given in units of~EeV.
	See the text for details.}

	\label{tab20151015}

	\medskip
	\centering
	\begin{tabular}{|l|c|c|c|c|}
		\hline
		Models & E & (E, dist, $\theta$, $\phi$) & E(*) & NL-NT E(*) \\
		\hline
		Pure mean$\pm$std & $1.48\pm0.25$ & $2.09\pm0.36$ & $2.47\pm0.16$
		                  & $2.62\pm0.23$ \\
		Pure median       & 1.45 & 2.07 & 2.46 & 2.62 \\
		Fixed mean$\pm$std
						      & $1.43\pm0.20$ & $1.99\pm0.29$ & $2.43\pm0.12$
						      & $2.59\pm0.21$ \\
		Fixed median      & 1.44 & 2.00 & 2.42 & 2.60 \\
		\hline
	\end{tabular}
\end{table}

While recognition of the track and reconstruction of energy of this
particular event might be improved by a more accurate procedure of
flat-fielding, we think its analysis raises at least two questions to be
studied in more detail. One of them relates to possible improvements of
the procedure of recognizing EAS tracks, especially on non-uniform
backgrounds. The other one calls for an in-depth study of the relevance
of models for energy reconstruction depending on the training data sets.
We plan to address these issues in the future studies.

\section{Discussion}

We have considered the task of reconstructing energy of three
ultra-high energy cosmic rays registered with a small fluorescence
telescope EUSO-TA, which operated at the site of the Telescope Array
experiment.  All events were recorded within a single time frame, which
does not allow applying the conventional procedure of energy
reconstruction.

Using several sets of simple artificial neural networks, we obtained
reasonable estimates of energy of two of the three UHECRs, close to
those reconstructed by the Telescope Array experiment with their much
larger and more sophisticated fluorescence telescopes.  Contrary to the
conventional approach, reconstruction of the shower geometry was not
needed to estimate energy of primary particles.  Energy estimates for
the two events were provided by models trained in two distinct ways: one
group of models was trained to reconstruct energy (and possibly distance
to the shower core and the UHECR arrival direction) using information
about the EAS track previously recognized by another ANN, while the
other could reconstruct energy without any prior knowledge about the
track. For one of the events, all types of models provided compatible
results, while for the other one estimates obtained with models trained
in a ``blind'' fashion occurred to be more accurate.  In other words, we
have found that reasonable estimates of energy can be obtained even
without recognizing the shower track on the photodetector.

The third of the events posed problems both with its track
recognition and estimation of energy of the UHECR. This does not allow
us to claim that the suggested method is universal and always works.  In
particular, we need to address open questions related to the dependence
of the accuracy of energy reconstruction on the quality of recognizing
EAS tracks (in models that need this). We also plan to consider other
approaches to obtaining estimations of energy, which do not need
training of a large number of models. Last but not least, it should be
studied how our estimations depend on the nature of primary particles
since we have only considered protons.  Nevertheless, we believe our
study demonstrates that neural networks suggest a promising approach for
reconstructing parameters of UHECRs registered with fluorescence
telescopes.

\bigskip

One of the authors, M.Z., wishes to thank Francesca Bisconti and
Zbigniew Plebaniak for numerous helpful discussions concerning EUSO-TA
and EUSO-Offline, and Antonio Giulio Coretti for insightful comments on
the manuscript.

The study was conducted under the state assignment of Lomonosov
Moscow State University and supported by Russian State Space Corporation
Roscosmos.
Search for EAS tracks in the PAIPS-L telescope data is being performed
with support from the Russian Science Foundation grant 22-62-00010.

\bibliography{eusota}

\end{document}